# The Effect of Magnetic Fields, Temperature and Current on the Resistivity of Bi-2223 High Temperature Superconductors

B. de Mayo

Department of Physics, University of West Georgia, Carrollton, GA 30118
bdemayo@westga.edu

The electrical resistivity of polycrystalline $Bi_2Sr_2Ca_2Cu_3O_{10+x}$ (Bi-2223) was measured vs. applied magnetic fields up to 0.45 T, applied currents up to 1 A, and temperature from liquid nitrogen temperature (LN2) to room temperature. In the lowest temperature region, the only truly zero resistivity was observed when the magnetic field was zero; otherwise, a quadratic dependence on the magnetic field occurred. Hysteresis was noted at the higher currents. Current vs. voltage curves in this region revealed a non-ohmic resistivity. In the transition region to the mixed state, indications of negative resistivity and suggestions of a phase change were observed. Arrhenius plots yielded activation energies of around 0.05 eV/molecule. In the mixed state region up to the transition temperature of ~110K, analysis implied that 4 superconducting quantum states exist and that they are cooperatively filled by the superconducting charge carriers. The occupation of the superconducting quantum states is negatively affected by the applied magnetic field and by the applied current. No effect on the polarity or direction of the magnetic field with respect to the direction of the current was observed.

Introduction

Even after 25 years of intensive experimental and theoretical work, a basic understanding of high temperature superconductivity in the cuprates [1] is elusive. In this study, we examined the effects of applied magnetic field and applied currents on the electrical resistivity of polycrystalline $Bi_2Sr_2Ca_2Cu_3O_{10+x}$ (Bi-2223) [2] at temperatures between liquid nitrogen (LN2) and room temperature.

Experimental

Four samples of optimally doped Bi-2223 were purchased from a commercial source, Colorado Superconductor Inc., in the form of polycrystalline discs of 2.5 cm in diameter and 0.25 cm thick. Each had 4 wires attached in a row; an electrical current was applied through the outer two leads; the resulting voltage was measured across the inner two contacts. A virtual instrument was devised using LabView, a product of National Instruments, Inc.; in a typical run, four measurements were made every half second: the elapsed time, the voltage, the magnetic field strength applied to the sample, and the resistance of a resistance temperature detector (RTD) (Omega Engineering model number 1PT100KN1510) which was in physical contact with the sample. For a typical experimental run, over 10,000 x 4 data points were usually recorded. The voltage and RTD values were measured using Keithley Instruments 2000 digital multimeters as part of the virtual instrument. Constant currents of up to 1.00 A were applied with a GW Instek programmable power supply PSP-603. The air cooled, iron core electromagnet, Quantum Electronics, Inc., Model B-2, was powered with an Alligent 6675A System VDC power supply. A double pole double throw knife switch in a crossover configuration allowed us to easily zero the field and to switch its polarity. The strength of the magnetic field was measured with a Daedalon magnetic flux density meter EP-15 with its probe attached to one face of the magnet.

The value of the field strength at the position of the sample (in the middle of the magnet gap) was found using an experimentally determined correction factor.

The resistivity of the disc-shaped samples was about 10% lower than samples with a rectangular cross-section of the same thickness and length, as determined experimentally. The cryostat was constructed of expanded styrene; the sample could be rotated 360° inside the cryostat while cooled. In a typical run, the sample was cooled to liquid nitrogen (LN2) temperature (which due to our altitude of around 305 m above sea level is lower than 77 K) and allowed to warm slowly.

Results

**I. The frozen vortex lattice region.** At LN2 temperature, the vortex lattice [3] is considered to be frozen. In this region, the resistivity was measured by the usual method of applying a current to the sample and measuring the voltage difference across two points. A magnetic field -0.45 T to +0.45 T, was applied parallel and perpendicularly to the round surface of the disc and parallel and perpendicular to the direction of the current. The resistivity was calculated to be the voltage divided by the current at each point. With zero magnetic field applied, the resistivity was always zero in this region. Fig. 1 shows the results for sample B2 with up to 1.0 A applied.

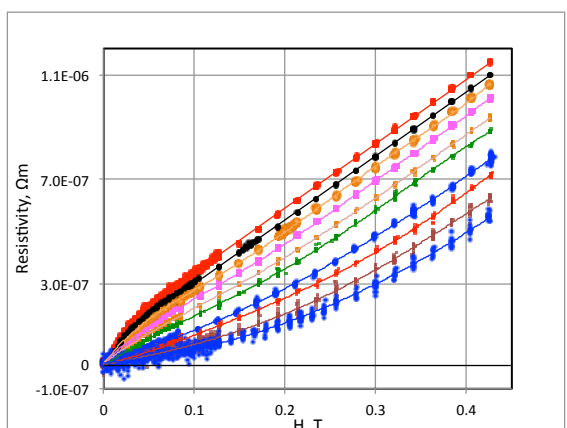

Figure 1. Resistivity at LN2 temperature vs. applied magnetic field, H, for 0.1A (bottom) to 1.0 A (top) (in intervals of 0.1A) applied to sample B2. The lines are guides for the eye.

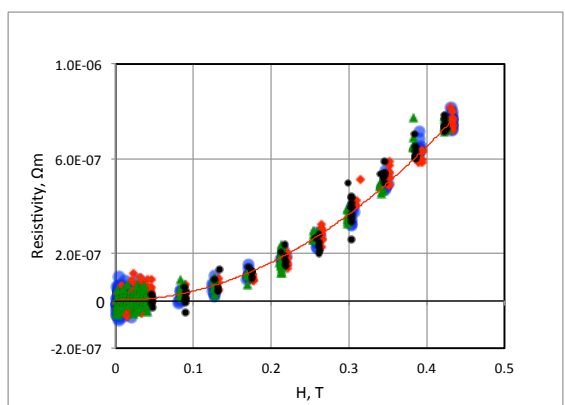

Figure 2. Resistivity vs. the magnitude of the applied magnetic field for 0.1A applied to sample B1. Symbols: blue, with the field parallel to the current and increasing in magnitude from zero; red, with the field in the same direction but decreasing in magnitude back to zero; green, increasing in magnitude in the opposite direction (anti-parallel to the current); and black, with the field still anti-parallel to the current but decreasing in magnitude back to zero. The red line is a quadratic fit to all of the data; see text for coefficients.

Notice that the curvature of the lines changes from positive to negative at around 0.8 A.

The resistivity showed little or no dependence on the polarity or the direction of the applied magnetic field with respect to the direction of the applied current. Fig. 2 shows the results when 0.1 A current was applied to sample B1 increasing from zero to 0.42 T, decreasing past zero to negative 0.42 T then increasing back to zero. The red line shows the quadratic fit to all of the data: Resistivity ($\Omega$m) = 4.35 x 10⁻⁶ ($\Omega$m/T²)H² -9.46 x 10⁻⁸ ($\Omega$m/T) H + 3.26 x 10⁻⁹ ($\Omega$m), where H is the applied magnetic field in Teslas and where the coefficient of determination $R^2$ = 0.982.

Fig. 3 compares the results for magnetic fields applied parallel and perpendicular to the current direction in sample B1; as can be seen there is little difference between the two. The curved lines are quadratic fits to the data: for the parallel case, Resistivity ($\Omega$m) = 2.77 x $10^{-6}$ ($\Omega$m/T$^2$) $H^2$ + 1.37 x $10^{-6}$ ($\Omega$m/T) H + 6.38 x $10^{-10}$ ($\Omega$m) with H in units of Teslas and with $R^2$ = 0.999. For the perpendicular case, Resistivity ($\Omega$m) = 2.38 x $10^{-6}$ ($\Omega$m/T$^2$) $H^2$ +1.33 x$10^{-6}$ ($\Omega$m/T) H -1.06 x $10^{-8}$ ($\Omega$m) with H in units of Teslas and with $R^2$ = 0.999. In Fig. 4, we see for 1.0 A applied to sample B1 that the upper part of the curve is linear, with Resistivity ($\Omega$m) = 2.68 x $10^{-6}$ ($\Omega$m/T) H (T) + 8.13 x $10^{-8}$ ($\Omega$m) with $R^2$ = 0.999. Fig. 5 shows that in the case of 1.0 A, the region of less than 0.05 T (500 Gs) for this sample the resistivity has some hysteresis. The lines are for quadratic fits to the data: for the increasing field from zero (blue) to 0.42 T, Resistivity ($\Omega$m) = -6.50 x $10^{-5}$ $H^2$ + 7.97 x $10^{-6}$ H -3.27 x $10^{-8}$, with H in Teslas and with $R^2$ =0.996, and for the field decreasing to zero (red) from 0.42 T, Resistivity ($\Omega$m) = 5.19 x $10^{-5}$ $H^2$ + 1.26 x $10^{-6}$ H - 3.58 x $10^{-9}$ with $R^2$ = 0.914. As before, H is the applied magnetic field in Teslas and $R^2$ is the coefficient of determination.

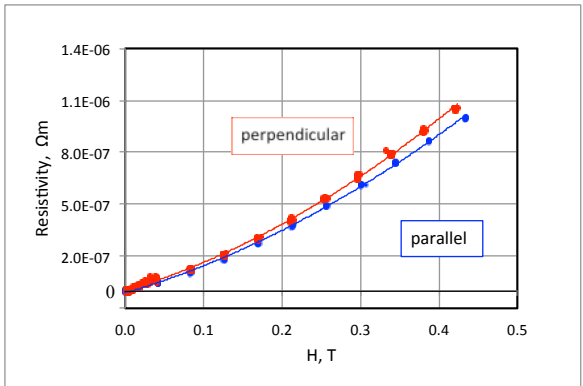

Figure 3. Resistivity vs. magnetic field applied parallel (blue) and perpendicularly (red) to the current, 0.4 A in this case. Sample B1. See text for description of curved lines.

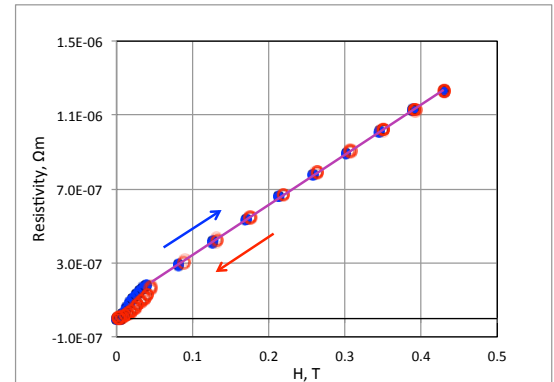

Figure 4. Resistivity for 1.0 A applied current vs. applied magnetic field, increasing from zero (blue) to 0.43 T and then decreasing to zero (red). Sample B1. See text for explanation of upper part straight line.

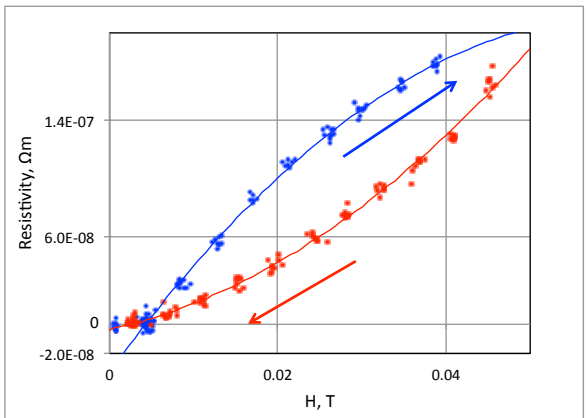

Figure 5. Lower portion of Fig. 4. Blue, magnetic field increasing from zero; red, magnetic field decreasing to zero from 0.43 T.

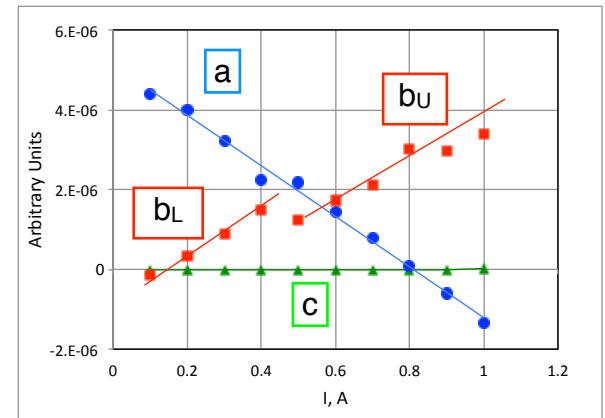

Figure 6. Coefficients of quadratic least squares fits for Fig. 1, a = coefficients of the square term; b = coefficients of the linear term; and c = the constant term. Units: for a, $\Omega$m/T$^2$; for b, $\Omega$m/T; and for c, $\Omega$m.

Fig. 6 shows the results of the quadratic fits of the data shown in Fig. 1; **a** (blue) is the coefficient of the quadratic term; as the current is increased, the curvature changes from positive to negative, crossing zero (linearity) at about 0.8 A. When the coefficient **a** is fit to a straight line we get: **a** ($\Omega$m/T$^2$) = -6.34 x 10$^{-6}$ ($\Omega$m/AT$^2$) I (A) + 5.13 x 10$^{-6}$ ($\Omega$m/T$^2$), with R$^2$ = 0.993. The slope of the linear term, **b** (red) appears to have a break at around 0.5 A; when the lower portion is fit to a straight line, we get $b_L$ ($\Omega$m/T) = 5.35 x 10$^{-6}$ ($\Omega$m/AT) I (A) - 7.01 x 10$^{-7}$ ($\Omega$m/T) with R$^2$ = 0.998. For the upper portion we get $b_U$ ($\Omega$m/T) = 4.40 x 10$^{-6}$ ($\Omega$m/AT) I (A) - 8.93 x 10$^{-7}$ ($\Omega$m/T) with R$^2$ = 0.951. A linear fit of the constant coefficient **c** (green) yields c ($\Omega$m) = 1.02 x 10$^{-8}$ ($\Omega$m/A) I (A) - 5.02 x 10$^{-9}$ ($\Omega$m) with R$^2$ = 0.181. Although these results are for one of the 4 samples used, B1, in every case the results for all three samples were very similar.

These measurements were taken at LN2 temperature by holding the applied current constant and varying the applied magnetic field, then repeating the process for different currents. However, the applied magnetic field was varied in the same way for each current. This allowed us to plot the average voltage for each field value vs. the applied current, as shown in Fig. 7. The resistance appears to be non-ohmic. The inset shows plots of the parameters of least squares fitting the individual curves to a quadratic function: as before, a, b, and c are the quadratic, linear and constant coefficients, respectively.

**II. Transition to the mixed state.** The next results were obtained by closely monitoring the time dependence of the measured parameters as heat leaked slowly into the cryostat. Fig. 8 shows the temperature (blue) and resistivity (red) vs. time for sample B3 with the applied field equal to zero. The resistivity is zero up to around 1250 s when it drops below zero and becomes negative. It then increases back to zero and dips again before rapidly increasing with the increasing temperature. Fig. 9 shows the results for different applied fields and Fig. 10 shows the results when a 3890 Gs = 0.3890 T field is applied. Fig. 11 shows the results for a different sample (B4) and another cryostat with an entirely different configuration. The same overall shape of the resistivity curve is preserved.

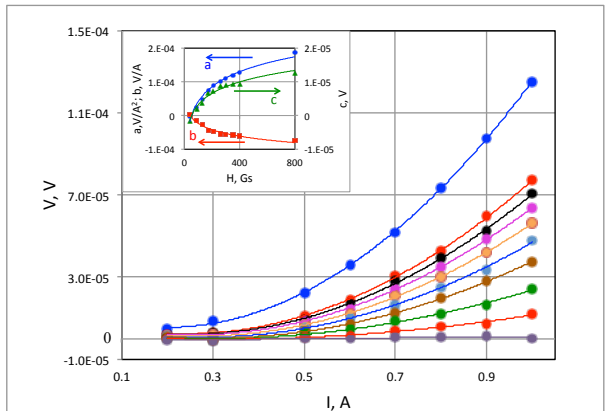

Figure 7. Voltage across the superconductor vs. the current applied at LN2 on sample B1 for various applied magnetic fields. Reading from the bottom up: 40, 90, 130, 170, 210, 300, 350, 400, and 800 Gs. Inset: coefficients for quadratic least squares fit to these curves: V = a x I$^2$ + b x I + c. The lines are aids for the eye. Sample B3.

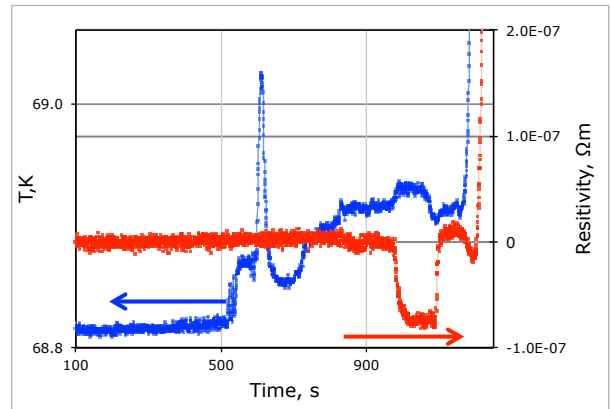

Figure 8. Temperature (blue) and resistivity (red) vs. time, sample B3. The applied magnetic field is zero.

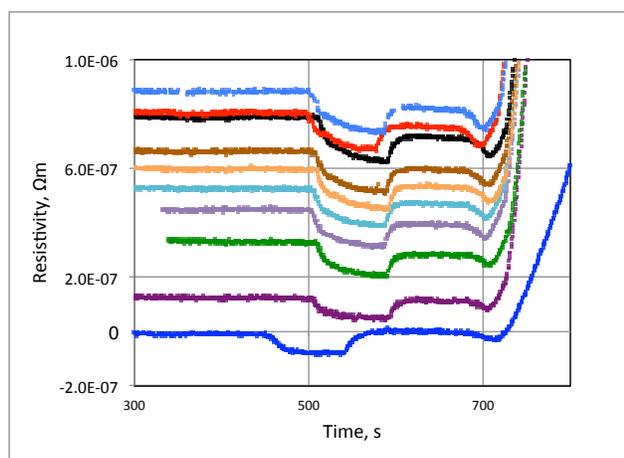

Figure 9. The resistivity vs. time for 10 different magnetic fields applied, all at 0.4 A applied current. From top to bottom, the fields were 376, 332, 290, 249, 207, 168, 123, 86, 43 and 0 Gs. The bottom results were from 4-4-2008 and the rest were from 4-08-2008, all from sample B3.

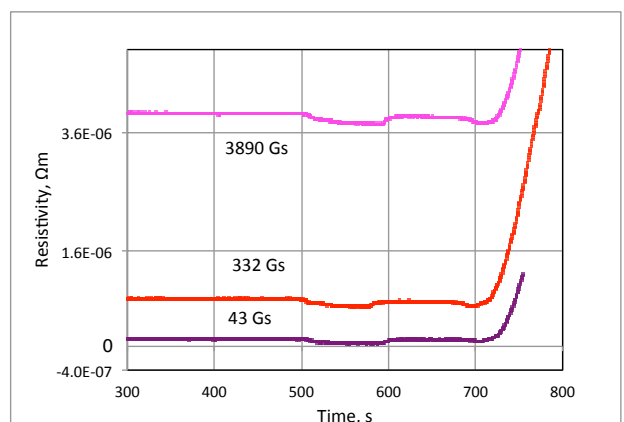

Figure 10. The resistivity vs. time for 3 different magnetic fields applied, all at 0.4 A applied current. From top to bottom, the fields were 3890, 332, and 43 Gs. Sample B3.

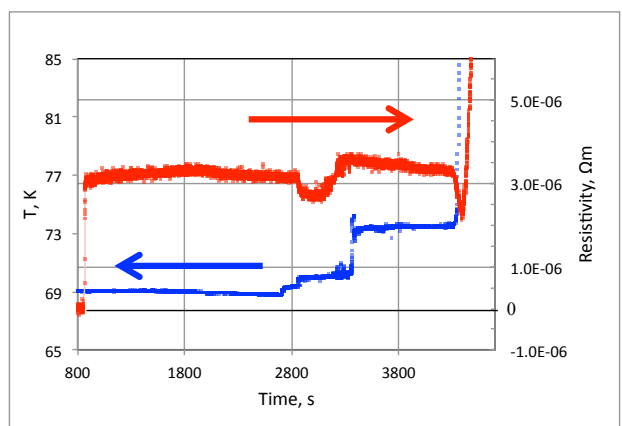

Figure 11. The temperature (blue) and resistivity (red) of sample B4 vs. time for an alternate cryostat, with a magnetic field of around 1000Gs applied by a small permanent magnet and with 0.4 A current applied.

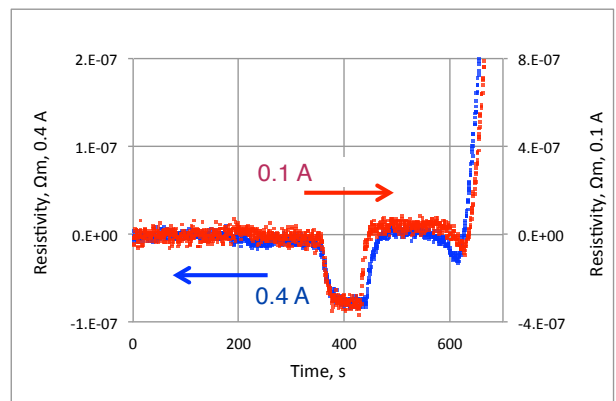

Figure 12. The resistivity vs. time with no magnetic field applied and with two different applied currents: 0.4 A (blue) and 0.1 A (red), for sample B3. Note the different vertical scales.

Fig. 12 gives the results of two different runs on the same sample with zero applied magnetic field, one with 0.4 A and the other with 0.1 A. The shapes are almost identical but the scatter is larger for 0.1 A applied current. Fig. 13 shows how the temperature (blue) is increasing with time even though the resistivity (red) is zero, then drops, then rises back to zero, then increases rapidly along with the temperature rise. Figure 14 shows the runs of Fig. 9 plotted vs. temperature for the range 69.0 to 70.0 K. The dots are all 0.5 seconds apart; notice the bunching up occurs at around 69.2 K for the 376 Gs line and at around 69.4 K for the zero field line, which indicates a slowing of the (slight) increase in resistivity with temperature. Fig. 15 and Fig. 16 show how the resistivity depends on the temperature for different applied magnetic fields and different applied currents, respectively.

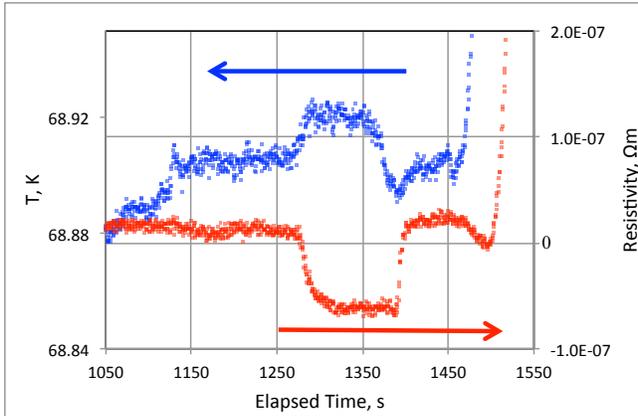

Figure 13. Closeup of Fig. 8. Temperature (blue) and resistivity (red) vs. time with 0 Gs and 0.4 A applied to sample B3.

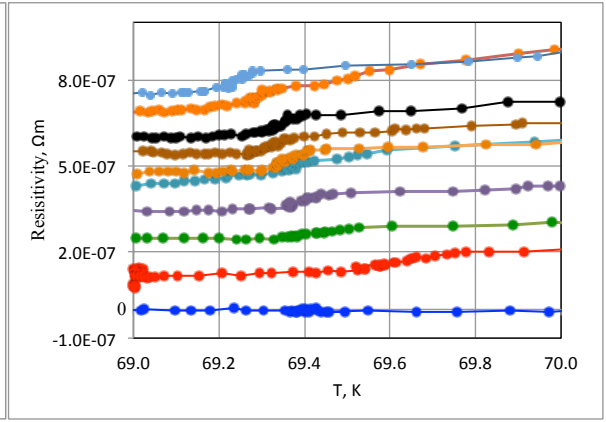

Figure 14. Resistivity for sample B3 with 0.4 A applied current and various magnetic fields, as in Fig. 9. From top to bottom, the fields were 3887, 331, 290, 249, 207, 164, 123, 86, 43 and 0 Gs.

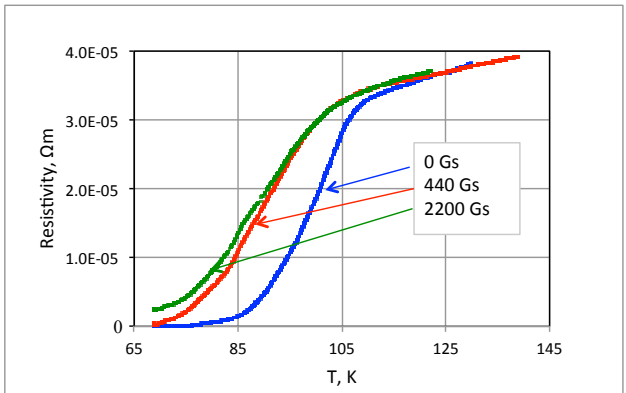

Figure 15. Resistivity vs. temperature with 3 applied magnetic fields and with 0.4 A applied current for sample B3.

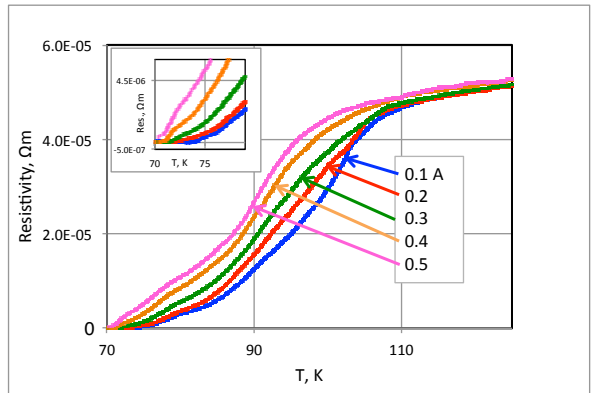

Figure 16. Resistivity vs. temperature for 5 different applied currents and zero magnetic field applied; sample B3. Inset: low temperature region.

**III. Arrhenius plots.** Using the well-known Arrhenius equation, the effect of temperature on the rate of a reaction can be shown. If the logarithm of a temperature dependent property, such as the resistivity in this case, is plotted as the ordinate against the inverse temperature as the abscissa, a straight line will result if the process is governed by a single rate-limited thermally activated process. The Arrhenius equation:

$$r = A\ \exp^{[-E_a/k_B T]},$$

becomes

$$\ln(r) = \ln(A) - [E_a/k_B T] = \ln(A) - [E_a/k_B]\ (1/T),$$

where $r$ is the resistivity, $T$ is the temperature, $A$ is the pre-exponential factor, $E_a$ is the activation energy per molecule, and $k_B$ is the Boltzmann constant.

Palstra, et al., [4] found for single crystal Bi$_{2.2}$Sr$_2$Ca$_{0.8}$Cu$_2$O$_{8+d}$ a current-independent, thermally activated resistance obeying Arrhenius' equation, with the activation energy weakly dependent on magnetic field.  They also found that the pre-exponential factor is 3 orders of magnitude larger than the normal state resistance and independent of the magnetic field and its orientation.  Zeldov [5] found a nonlinear current dependence of the activation energy in Bi$_2$Sr$_2$CaCu$_2$O$_8$ epitaxial films and Briceño [6] found that for single crystal Bi$_2$Sr$_2$CaCu$_2$O$_8$ the activation energy is temperature dependent in the 25K to 100K range.  The Arrhenius plot of the data of Safar, et al., [7] on single crystal BSCCO samples did not yield a single constant value for the activation energy, but instead implied that there are different activation energies for different conditions of temperature and applied magnetic fields.  Their results seemed to validate the vortex glass model [3].

Arrhenius plots of some of our data are shown in Fig. 17 and Fig. 18.  Fig. 17 shows the Arrhenius plot for sample B3 with 0.2 A applied current and fields of 210 and 2024 Gs applied.  Fig. 18 shows the result for sample B3 with zero magnetic field applied and currents of 0.1 A and 0.5A applied to the sample.  For Fig. 17, the slopes of the regions AB, BE, AC and CD, yield activation energies of 0.035, 0.070, 0.027, and 0.053 eV/molecule, respectively.  For Fig. 18, the straight lines show the linear fit to the data for each linear region and yield activation energies of 0.078 and 0.056 eV/molecule for 0.1 A and 0.5 A, respectively. The sharp drop offs for the zero magnetic field cases (Fig. 18) are reflections of the fact that, with zero applied magnetic field, the resistivity is absolutely zero, see Fig. 8-13.

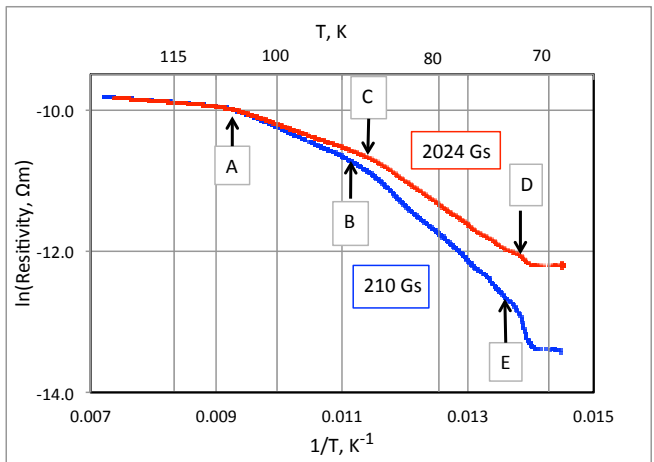

Figure 17. Arrhenius plot, sample B3, 0.2 A applied current, H = 2024 Gs (red) and H = 210 Gs (blue). The slopes of the regions AB, BE, AC and CD, yield activation energies of 0.035, 0.070, 0.027, and 0.053 eV/molecule, respectively.

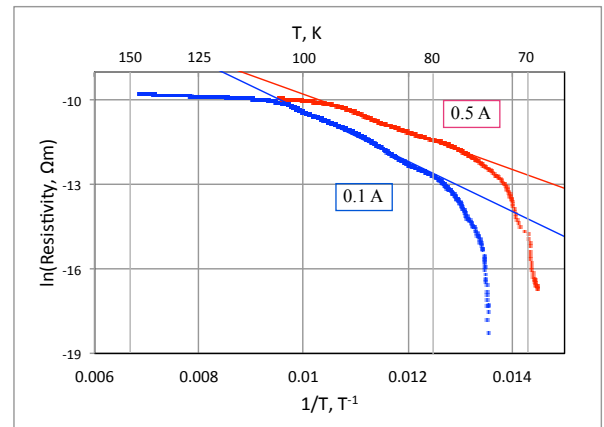

Figure 18. Arrhenius plot, sample B3, applied current of 0.1 A (blue) and 0.5 A (red), and H = 0 for both.  The straight lines show the linear fit to the data for each linear region and yield activation energies of 0.078 and 0.056 eV/molecule for 0.1 A and 0.5 A, respectively.

## IV. Above 115 K.

As the temperature rises, the resistivity increases in a manner observed by Subramanian et al. for bismuth high temperature superconductors in 1988 [2].  A typical run for our samples is shown in Fig. 19: the <u>blue</u> region shows the approach to normal behavior (from below); above the <u>red</u> transition region, the resistivity increases linearly with temperature (<u>green</u>) at a rate for this case of $1.08 \times 10^{-7}$ $\Omega$m/K, or $4.1 \times 10^{-5}$ $\Omega$/K. This can be compared to $2.7 \times 10^{-4}$ $\Omega$/K obtained from the data of reference 2 for polycrystalline $Bi_2Sr_{3-x}Ca_xCu_2O_{8+y}$ .

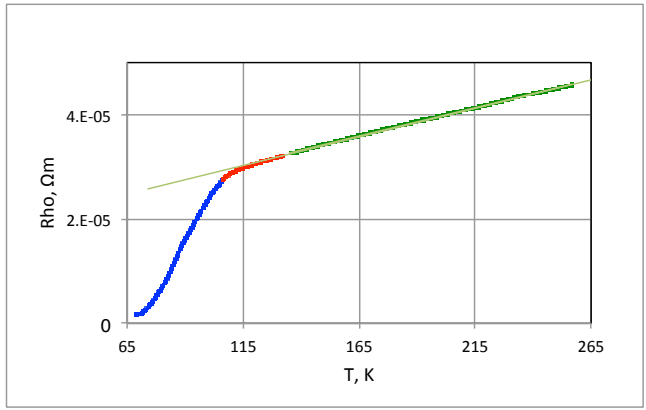

Figure 19. Resistivity for sample B3 with 0.3 A applied current and 408 Gs applied magnetic field. See text for a description of the colored regions.

## V. The Hill Equation.

The mixed state behavior can be analyzed using the Hill equation, which was originally introduced by Archibald Hill in 1910 to describe the oxygen binding of hemoglobin [8]. According to Hill, the fraction of hemoglobin with bound oxygen (the fractional degree of saturation), Y, is given as

$$Y(p) = \frac{p^n}{p^n + p_{1/2}}$$

where p is the partial pressure of oxygen, n is defined as the Hill parameter, and $p_{1/2}$ is the partial pressure for 50% saturation. The Hill parameter, n, gives the cooperativeness of the binding: n = 1 indicates independent or non-cooperative binding--filling of available binding sites proceeds independently of whether any other sites are filled. When n > 1, cooperative binding occurs, but n can't be higher than the number of binding sites. For hemoglobin, n is around 2.5 and the number of binding sites is 4. The parameter $p_{1/2}$ gives an indication as to how the binding is proceeding as the partial pressure (amount of oxygen available) increases. A low $p_{1/2}$ says that the filling proceeds well. A higher $p_{1/2}$ means that more oxygen is needed for an equivalent binding.

In 1904 Christian Bohr (Niels' father) noted the effect of acidity (pH) on the oxygen binding, Fig. 20 [9]. Increasing the pH of the solution containing the hemoglobin decreases the ability of oxygen atoms to attach to the hemoglobin sites: more oxygen is needed for an equivalent binding.

In our case, if we change axes and plot a "reverse resistivity" $R_r$ versus a "reverse temperature" $T_r$ :

$$R_r = 1 - (R - R_{Tc2})/(R_{Tc1} - R_{Tc2})$$

$$T_r = (T_{c1} - T)/(T_{c1} - T_{c2}),$$

where R is the resistivity, $R_{Tc1}$ is the resistivity at the superconductivity upper transition temperature $T_{c1}$ (~110 K), and $R_{Tc2}$ is the resistivity at the temperature at which the vortex lattice "un-freezes" $T_{c2}$ (~70 K). Then our Fig. 15, becomes Fig. 21, and Fig. 16 becomes Fig. 22. When T = $T_{c1}$, $T_r$ is zero and the number of available potential superconducting charge carriers is zero also; the reverse resistivity is zero, too, and the sample is in the normal state. When T = $T_{c2}$, $T_r$ = 1 and the sample is in the purely superconducting state ($R_r$ = 1), if the applied magnetic field is zero.

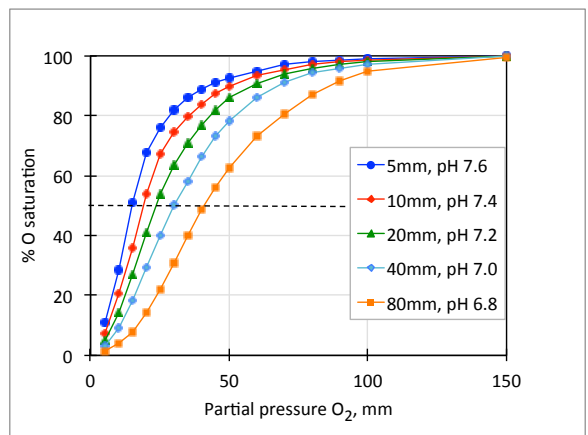

Figure 20. Bohr-Hill plot: percent oxygen saturation vs. partial pressure of oxygen for various partial pressures (pH) of carbon dioxide, [9].

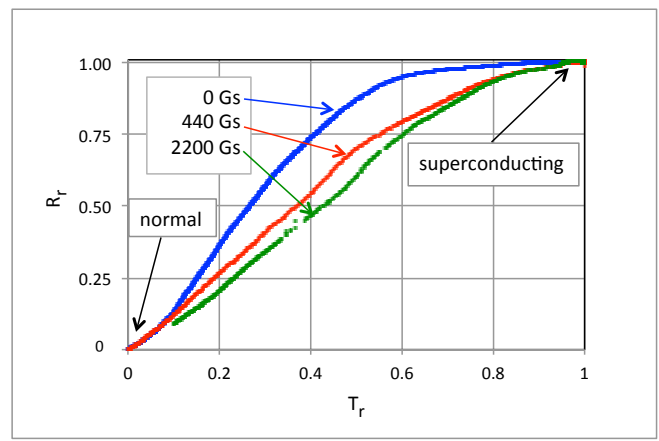

Figure 21. Hill plot, reduced resistivity vs. reduced temperature for sample B3 with 3 different applied magnetic fields, and with as in Fig. 15.

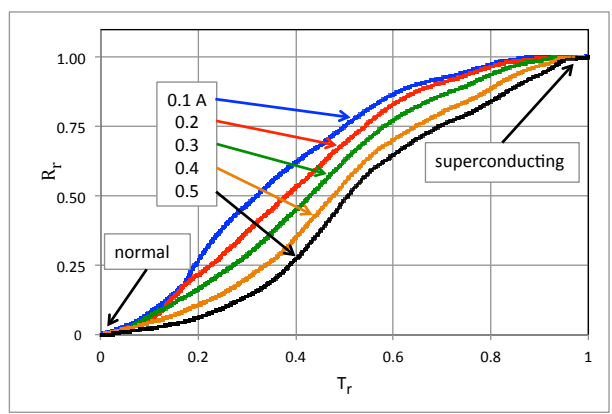

Figure 22. Reverse resistivity, $R_r$, vs. reverse temperature, $T_r$, for sample B3 with 5 different applied currents and with 408 Gs applied field, as in Fig. 16.

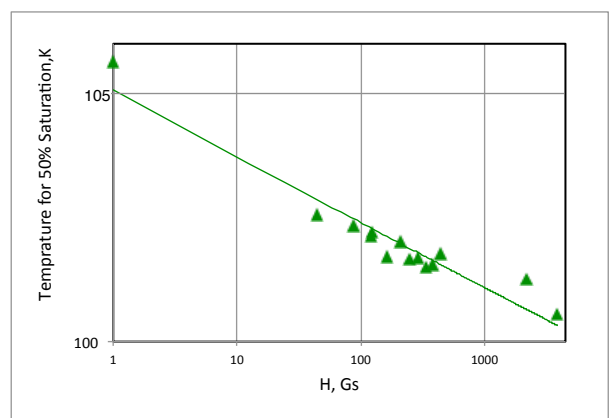

Figure 23. Log-log plot of the temperature for 50% occupancy, $T_{50}$, vs. the applied magnetic field, H, for sample B3 with 0.4 A applied current. The straight line is a power function fit to the data; see text for coefficients.

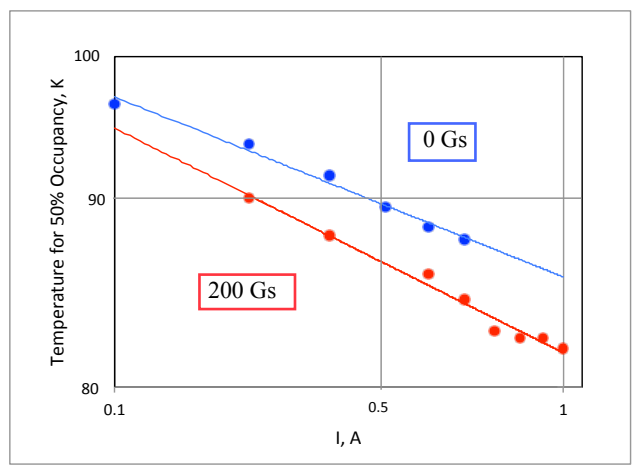

Figure 24. Log-log plot of the temperature for 50% occupancy, $T_{50}$, vs. the applied current, I, for sample B3 at zero applied magnetic field (blue) and with 200 Gs applied (red). The lines represent power function fits to the data; see text for coefficients.

For all runs, the n-value or rate was relatively constant at about $2.8 \pm 0.2$ indicating coop­erative filling of 4 quantum states or receptor sites. Fig. 23 shows a log-log plot of the tempera­ture for 50% occupancy, $T_{50}$, vs. the applied magnetic field for sample B3 with 0.4A current ap­plied. The straight line is a power function fit to the data: $T_{50} = a \, H^b$, units of Kelvin for the temperature and Gauss for the magnetic field; here $a = 105.1$ and with $b = -5.586 \times 10^{-3}$ and $R^2 = 0.928$. Fig. 24 shows the temperature for 50% occupancy, $T_{50}$, vs. the applied current, I, for sample B3 with zero magnetic field applied (blue) and 200 Gs applied (red). The temperature is in units of Kelvin and the current is in units of Amps. The lines represent power function fits to the data: $T_{50} = a \times I^b$, with units of Kelvin for T and Amps for I. For zero applied field (blue), a $= 86.17$ and $b = -5.282 \times 10^{-2}$ with $R^2 = 0.985$ and for 200 Gs applied (red), $a = 81.86$, $b = -6.592$ $\times 10^{-2}$, and $R^2 = 0.982$.

These results can be heuristically explained in part with the following: In zero magnetic field, when $T = T_{c1}$, the reverse temperature $T_r$ is zero and the number of available potential su­perconducting charge carriers is zero also; the reverse resistivity is zero, too, and the sample is in the normal state. As the temperature drops, $T_r$ increases, as does the the number of available po­tential superconducting charge carriers. These begin to cooperatively occupy the 4 different su­perconducting quantum states. At $T = T_{c2}$, the number of available superconducting charge carri­ers is a maximum (in zero field), as is the occupation of the superconducting quantum states, and the reverse resistivity levels off. If the applied field is zero, the reverse resistivity is one and the resistivity is exactly zero. For values of the applied magnetic field greater than zero, the oc­cupation by the available potential superconducting charge carriers is inhibited by the magnitude (but not the direction) of the applied magnetic field, and the resistivity has a finite value, even at temperatures less than $T_{c2}$. The applied current also has an inhibitive effect on the occupation of the superconducting quantum states, as can be seen in Fig. 16. However, note that while any magnetic field results in a finite resistivity, even at temperatures less than $T_{c2}$, this is not the case for the current (Fig. 16), which shows that the temperature for 50% occupancy, $T_{50}$, is higher for the lower currents. In other words, it is easier to fill the superconducting quantum states if the applied current is lower.

Conclusions: The effect of applied current, temperature and applied magnetic field on the resis­tivity of polycrystalline B-2223 was measured between the temperatures of LN2 and the super­conducting transition temperature of ~110 K. There seems to be a frozen vortex lattice even when the applied magnetic field is zero. There are indications of negative resistivity in the region of the melting of the vortex lattice. Voltage vs. current plots show non-Ohmic behavior. Ar­rhenius plots imply activation energies of around $0.05 \pm 0.02$ eV/molecule. The Hill equation was used to conclude that there are probably 4 superconducting quantum states in this substance, and that an applied magnetic field and/or an applied current inhibits the filling of these states. Also, the temperature seems to determine the number of charge carriers that are suitable for fill­ing these states.

Acknowledgements: The author gratefully acknowledges the generous support of the Georgia Space Grant Consortium-NASA NNG05GJ65H and the helpful assistance of numerous undergraduate research workers.

References

[1] Bednorz, J. G., and K. A. Mueller, Possible High $T_C$ Superconductivity in the Ba-La-Cu-O System, Zeitschrift für Physik B **64** (2), 189–193 (1986).

[2] Subramanian, M. A., C. C. Torardi, J. C. Calabrese, J. Gopalakrishnan, K. J. Morressey, T. R. Askew, R. B. Flippen, U. Chowdhry,and A. W. Sleight, A New High-Temperature Superconductor: $Bi_2Sr_{3-x}Ca_xCu_2O_{8+y}$, Science **239** (4843), 1015-1017 (26 February 1988).

[3] Abrikosov, A. A., On the Magnetic Properties of Superconductors of the Second Group, Soviet Physics JETP **5** (6), 1174-1182 (1957).

[4] Palstra, T. T. M. Palstra, B. Batlogg, L. F. Schneemeyer, and J. V. Waszczak, Thermally Activated Dissipation in $Bi_{2.2}Sr_2Ca_{0.8}Cu_2O_{8+\delta}$, Phys. Rev. Lett. **61** (14), 1662-1665 (3 October 1988).

[5] Zeldov, E., N. M. Amer, G. Koren, and A. Gupta, Flux Creep in $Bi_2Sr_2CaCu_2O_8$ Epitaxial Films, Appl. Phys. Lett. **56** (17), 1700-1702 (23 April 1990).

[6] Briceño, M. F. Crommie, and A. Zettl, Giant Out-of-Plane Manetoresistance in Bi-Sr-Ca-Cu-O: A New Dissipatation Mechanism in Copper-Oxide Superconductors?, Phys. Rev. Lett. **66** (16), 2164-2166 (22 April 1991).

[7] Safar, H., P. L. Gammel, D. J. Bishop,D. B. Mitzi and A. Kapitulnik, SQUID Picovoltometry of Single Crystal $Bi_2Sr_2CaCu_2O_{8+\delta}$: Observation of the Crossover from High-Temperature Arrhenius to Low-Temperature Vortex-Glass Behavior, Phys. Rev. Lett. **68**, 2672–2675 (27 April 1992).

[8] Hill, A. V., The Possible Effects of the Aggregation of the Molecules of Hæmoglobin on its Dissociation Curves, J. Physiol. Suppl. **40**, iv-vii (January 22, 1910).

[9] Bohr, C., K. Hasselbalch, and A. Krogh, Concerning a Biologically Important Relationship - The Influence of the Carbon Dioxide Content of Blood on its Oxygen Binding, Skand. Arch. Physiol. **16**, 401-412 (1904).